\documentstyle[multicol,aps,psfig]{revtex}
\renewcommand{\narrowtext}{\begin{multicols}{2}
\global\columnwidth20.5pc\noindent}
\renewcommand{\widetext}{\end{multicols}
\global\columnwidth42.5pc}
\begin{document}
\draft
\preprint{May 1999}
\title{Critical Behavior of Anisotropic Heisenberg Mixed-Spin Chains
       in a Field}
\author{T$\hat{\mbox o}$ru Sakai}
\address
{Faculty of Science, Himeji Institute of Technology,
 Ako, Hyogo 678-1297, Japan}
\author{Shoji Yamamoto}
\address
{Department of Physics, Okayama University,
 Tsushima, Okayama 700-8530, Japan}
\date{January 6, 1999}
\maketitle
\begin{abstract}
We numerically investigate the critical behavior of the
spin-$(1,\frac{1}{2})$ Heisenberg ferrimagnet with anisotropic
exchange coupling in a magnetic field.
A quantized magnetization plateau as a function of the field,
appearing at a third of the saturated magnetization, is stable over
whole the antiferromagnetic coupling region.
The plateau vanishes in the ferromagnetic coupling region via the
Kosterlitz-Thouless transition.
Comparing the quantum and classical magnetization curves, we elucidate
what are essential quantum effects.

\end{abstract}
\narrowtext

   Quantized plateaux in magnetization curves as functions of a
magnetic field for spin chains have been attracting much current
interest.
The trimerized spin-$\frac{1}{2}$ chain exhibits a massive phase at
$m/m_{\rm sat}=\frac{1}{3}$ \cite{Hida59,Okam45}, while the dimerized
spin-$1$ chain at $m/m_{\rm sat}=\frac{1}{2}$ \cite{Tone17,Tots03},
where $m$ is the magnetization per unit period and $m_{\rm sat}$ is
its saturation value.
The presence of finite gap and plateau has further been discussed
\cite{Naka26,Tots54,Cabr,Cabr26,Kole} and actually been observed
\cite{Naru09,Shir48} for various polymerized spin chains and ladders.
It may be the Lieb-Schultz-Mattis theorem \cite{Lieb07} and its
generalization \cite{Tots03,Affl86,Oshi84} in recent years that
motivate such vigorous arguments.
Oshikawa, Yamanaka, and Affleck (OYA) pointed out that quantized
plateaux in magnetization curves may appear satisfying the condition
\begin{equation}
   \widetilde{S}-m=\mbox{integer}\,,
   \label{E:OYA}
\end{equation}
where $\widetilde{S}$ is the sum of
spins over all sites in the unit period.

   The OYA argument stimulates us to study quantum mixed-spin chains
as well.
An arbitrary alignment of alternating spins $S$ and $s$ in a magnetic
field, which is described by the Hamiltonian
\begin{equation}
   {\cal H}
    ={\displaystyle\sum_{j=1}^N}
    \left[
     (\mbox{\boldmath$S$}_{j}\cdot\mbox{\boldmath$s$}_{j})_\alpha
    +(\mbox{\boldmath$s$}_{j}\cdot\mbox{\boldmath$S$}_{j+1})_\alpha
    -H(S_j^z+s_j^z)
    \right]\,,
   \label{E:H}
\end{equation}
with
$(\mbox{\boldmath$S$}\cdot\mbox{\boldmath$s$})_\alpha
 =S^xs^x+S^ys^y+\alpha S^zs^z$,
shows ferrimagnetism \cite{Lieb49}, instead of antiferromagnetism,
and is another current topic from both theoretical
\cite
{Dril13,Alca67,Pati94,Breh21,Nigg31,Yama09,Ono76,Kura62,Yama08,Ivan24,Mais08,Yama}
and experimental
\cite{Kahn91,Hagi09}
points of view.
As $H$ increases from zero to the saturation field
\begin{equation}
   H_{\rm sat}
    =\alpha(S+s)+\sqrt{\alpha^2(S-s)^2+4Ss}\,,
   \label{E:Hsat}
\end{equation}
the OYA criterion (\ref{E:OYA}) allows us to expect quantized plateaux
at $m=S+s-1$, $S+s-2$, $\cdots$, $1$ (or $\frac{1}{2}$).
Since the low-energy physics of the model (\ref{E:H}) is qualitatively
the same \cite{Alca67,Pati94,Yama} regardless of $S$ and $s$ as long
as $S\neq s$, here, let us consider the simplest case
$(S,s)=(1,\frac{1}{2})$.
Then a plateau may appear at $m=\frac{1}{2}$.
At the Heisenberg point, the ground state of the Hamiltonian
(\ref{E:H}) without field is a multiplet of spin $(S-s)N$
\cite{Lieb49} and thus has elementary excitations of two distinct
types \cite{Pati94,Breh21,Yama09}.
The ferromagnetic excitations, reducing the ground-state
magnetization, exhibit a gapless dispersion relation, whereas the
antiferromagnetic ones, enhancing the ground-state magnetization,
are gapped from the ground state.
Therefore, at the isotropic point, $m$ as a function of $H$ should
jump up to $\frac{1}{2}$ just as the field is applied and remain
unchanged until the field reaches the antiferromagnetic excitation
gap $1.759$ \cite{Yama09}.
\begin{figure}
\mbox{\psfig{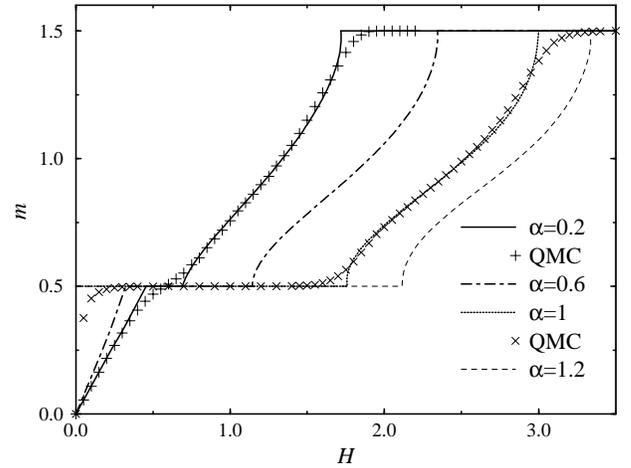}}
\vskip 5mm
\caption{The ground-state magnetization curves for the quantum
         Hamiltonian (2) at $\alpha>0$, where QMC calculations at
         $N=32$ and $T=0.08$ are also shown for comparison.}
\label{F:mAF}
\end{figure}
\vskip 3mm

   Once we turn on the exchange-coupling anisotropy, the plateau is
not so trivial any more.
We show in Fig. \ref{F:mAF} the zero-temperature magnetization curves
of the anisotropic chains, which have been calculated by the
numerical diagonalization technique combined with a
finite-size scaling 
analysis \cite{Saka83}.
In order to verify the reliability of our scaling analysis, which is
briefly explained later, we have carried out quantum Monte Carlo
(QMC) calculations \cite{Yama49} as well at $N=32$ and $N=16$, where,
due to the small correlation length \cite{Pati94,Breh21} of the
system, the data show no size dependence beyond the numerical
uncertainty.
Although the QMC findings are obtained at a sufficiently low but
finite temperature, they fully suggest that the present
diagonalization-based calculations well describe the
thermodynamic-limit properties.
As the model approaches the Ising limit ($\alpha\rightarrow\infty$),
the plateau monotonically grows and ends up with a stepwise
magnetization curve.
On the other hand, the introduction of the $XY$-like coupling
anisotropy reduces the plateau.
Thus we take great interest in where and how the plateau vanishes.
Alcaraz and Malvezzi \cite{Alca67} showed that the ground state of
the model (\ref{E:H}) without field is in the critical phase over the
whole region $-1\leq\alpha<1$.
At the Heisenberg point, the model still lies in the massless phase
but the low-energy dispersion as a function of momentum is
quadratic \cite{Yama09}.
For $\alpha>1$, the model is in the massive phase \cite{Ono76} and
its low-energy structure is well understood by the spin-wave
dispersions
\begin{equation}
   \omega_k^\mp
    =\sqrt{\alpha^2(S+s)^2-4Ss\cos^2\mbox{$\frac{k}{2}$}}
    \mp\alpha(S-s)\,,
\end{equation}
which describe the sector of the magnetization
$\sum_j(S_j^z+s_j^z)\equiv M<(S-s)N$ and that of $M>(S-s)N$,
respectively.
Thus the introduction of the anisotropy essentially changes the
nature of the model (\ref{E:H}) and a fascinating physics must lie
especially in the $XY$-like coupling region.
In the present article, we clarify how the quantized plateau of the
$(1,\frac{1}{2})$ model behaves as a function of the anisotropy and
aim to reveal critical phenomena inherent in quantum ferrimagnets.

   In order to investigate the quantum critical behavior, we carry out
a scaling analysis on the numerically calculated energy spectra of
finite clusters up to $N=12$.
Let $E(N,M)$ denote the lowest energy in the subspace with a fixed
magnetization $M$ for the Hamiltonian (\ref{E:H}) without the Zeeman
term.   
The upper and lower bounds of the field which induces the ground-state
magnetization $M$ are, respectively, given by
\begin{eqnarray}
   &&
   H_+(N,M)=E(N,M+1)-E(N,M),
   \label{E:H+}
   \\
   &&
   H_-(N,M)=E(N,M)-E(N,M-1).
   \label{E:H-}
\end{eqnarray}
If the system is massive at the sector labeled $M$, $H_\pm(N,M)$
should exhibit exponential size corrections and result in different
thermodynamic-limit values $H_\pm(m)$, which can precisely be
estimated through the Shanks' extrapolation \cite{Shan01}.
For the critical system, on the other hand, $H_\pm(N,M)$ are expected
to converge to the same value as \cite{Card85}
\begin{equation}
   H_{\pm}(N,M)
    \sim H(m) \pm \frac{\pi v_{\rm s}\eta}{N}\,,
    \label{E:CFT}
\end{equation}
where $v_{\rm s}$ is the sound velocity and $\eta$ is the critical
index defined as 
$\langle\sigma^+_0\sigma^-_r\rangle \sim (-1)^r r^{-\eta}$
for the relevant spin operator $\sigma$, which may here be an
effective combination of $\mbox{\boldmath$S$}$ and
$\mbox{\boldmath$s$}$. 

\begin{figure}
\mbox{\psfig{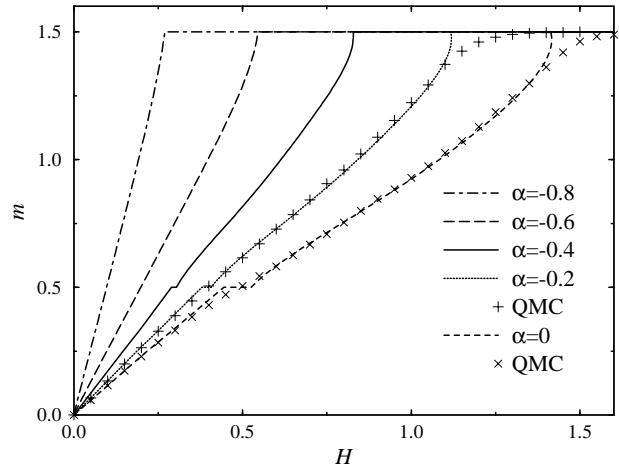}}
\vskip 5mm
\caption{The same as Fig. 1 but at $\alpha\leq 0$.}
\label{F:mF}
\end{figure}

   Figure \ref{F:mAF} was thus obtained, where we smoothly
interpolated the raw data $H(m)$.
The system is trivially gapless at all the sectors of $M$
for $\alpha\leq -1$ and should therefore encounter a massive-massles
phase transition in the $XY$-like coupling region.
Now we present in Fig. \ref{F:mF} the magnetization curves at
$\alpha\leq 0$ so as to detect the transition.
Surprisingly, the plateau still exists at the $XY$ point ($\alpha=0$)
and the transition occurs in the ferromagnetic-coupling region.
At an naive idea of relating the massive state with the staggered
N\'eel-like order in the direction of the external field, we are
never able to understand why the plateau is so stable against the
$XY$-like anisotropy.
\begin{figure}
\vskip -5mm
\mbox{\psfig{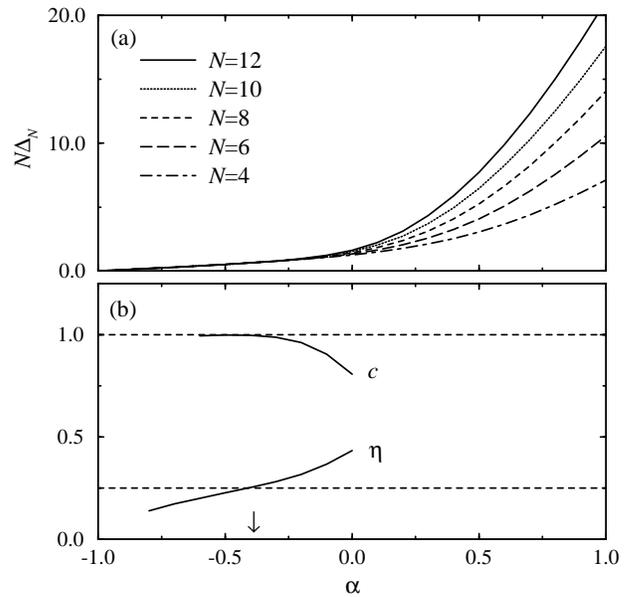}}
\vskip 5mm
\caption{(a) Scaled quantity $N{\mit\Delta}_N$ versus $\alpha$.
         (b) The central charge $c$ and the critical exponent $\eta$
         versus $\alpha$ in the vicinity of the phase boundary.}
\label{F:CP}
\end{figure}
\begin{figure}
\vskip -5mm
\mbox{\psfig{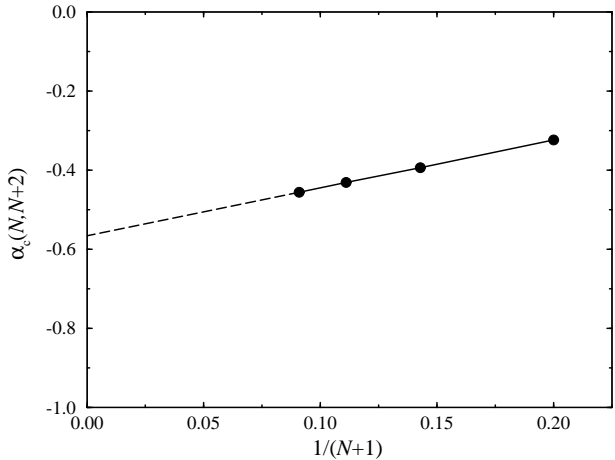}}
\vskip 5mm
\caption{The linear extrapolation of the size-dependent fixed point
         $\alpha_{\rm c}(N,N+2)$.}
\label{F:PRG}
\end{figure}

   The plateau length ${\mit\Delta}_N=H_+(N,M)-H_-(N,M)$ is a relevant
order parameter to detect the phase boundary. 
When the system is critical, ${\mit\Delta}_N$ should be proportional
to $1/N$ because of the scaling relation (\ref{E:CFT}).
We plot in Fig. \ref{F:CP}(a) the scaled quantity $N{\mit\Delta}_N$
as a function of $\alpha$.
$N{\mit\Delta}_N$ is almost independent of $N$ in a finite range of
$\alpha$, rather than at a point, which exhibits an aspect of the
Kosterlitz-Thouless(KT) transition \cite{Kost81}.
It is likely that the $XY$-like anisotropy induces the KT transition
\cite{Soly80} followed by the gapless spin-fluid phase whose spin
correlation shows a power-law decay.
Let us evaluate the central charge $c$ of the critical phase, which
is expected to be unity.
The asymptotic form of the ground-state energy \cite{Card85}
\begin{equation}
   \frac{E(N,M)}{N}
    \sim \varepsilon(m)-\frac{\pi cv_{\rm s}}{N^2}\,,
\label{E:GS}
\end{equation}
allows us to extract $c$ from the finite-cluster energy spectrum
provided $v_{\rm s}$ is given.
Here we calculate $v_{\rm s}$ as
\begin{equation}
   v_{\rm s}
    =\frac{N}{2\pi}
     \left[
      E_{k_1}(N,M)-E(N,M)
     \right]\,,
\label{E:vs}
\end{equation}
where $k_1=2\pi/N$ and $E_k(N,M)$ is the lowest energy in the
subspace specified by the momentum $k$ as well as by the magnetization
$M$.
The size correction for the formula (\ref{E:vs}) is of order
$O(1/N^2)$, which is essentially negligible in the present system.
In Fig. \ref{F:CP}(b) we plot $c$ versus $\alpha$ and find that $c$
approaches unity as the system goes toward the critical region.
We further investigate the critical exponent $\eta$ so as to verify
the KT universality and to specify the phase boundary.
In the critical region the asymptotic formula
${\mit\Delta}_N\sim 2\pi v_{\rm s}\eta/N$
enables us to estimate $\eta$.
Since the KT transition holds $\eta=\frac{1}{4}$ at the phase
boundary, we can evaluate the transition point $\alpha_{\rm c}$ from
$\eta$ as a function of $\alpha$.
Figure \ref{F:CP}(b) claims that $\alpha_{\rm c}=-0.41\pm 0.01$,
where $c=1.00\pm 0.01$.
The phenomenological renormalization-group (PRG) technique
\cite{Nigh61} is another numerical tool to determine the phase
boundary.
Taking ${\mit\Delta}_N$ as the order parameter, we extract the
size-dependent fixed point $\alpha_{\rm c}(N,N+2)$ from the PRG
equation
\begin{equation}
   (N+2){\mit\Delta}_{N+2}(\alpha)
    =N{\mit\Delta}_N(\alpha )\,. 
\label{prg}
\end{equation}
In Fig. \ref{F:PRG} we plot $\alpha_{\rm c}(N,N+2)$ as a function
of $1/(N+1)$, which is linearly extrapolated to $\alpha_{\rm c}=-0.57$.
The PRG estimate is somewhat discrepant from the above-obtained phase
boundary.
Here we should be reminded of Nomura-Okamoto's enlightening analysis
\cite{Nomu23} on usage of the PRG method.
The PRG equation applied to a gapful-gapful phase transition yields
a reliable estimate of the critical point, whereas, for a transition of
KT type, the PRG estimate is quite likely to miss the exact solution
due to the incidental logarithmic size correction, encroaching upon
the KT-phase region. 
Considering the limited availability of the PRG analysis, we may
recognize the present PRG solution as the lower boundary of
$\alpha_{\rm c}$.
\begin{figure}
\mbox{\psfig{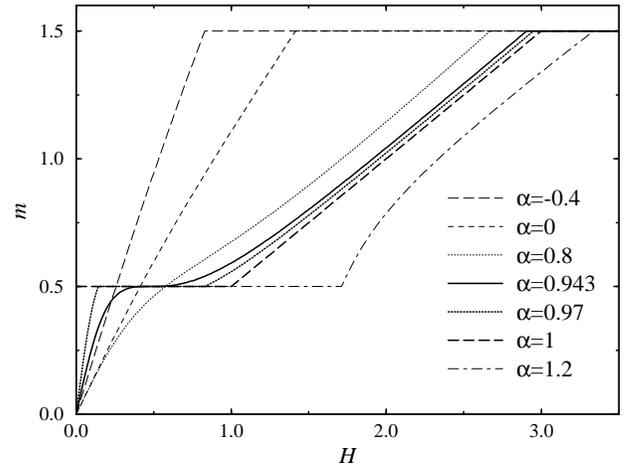}}
\vskip 3mm
\caption{The ground-state magnetization curves for the classical
         Hamiltonian (2).}
\label{F:mclass}
\end{figure}
\begin{figure}
\vskip -5mm
\mbox{\psfig{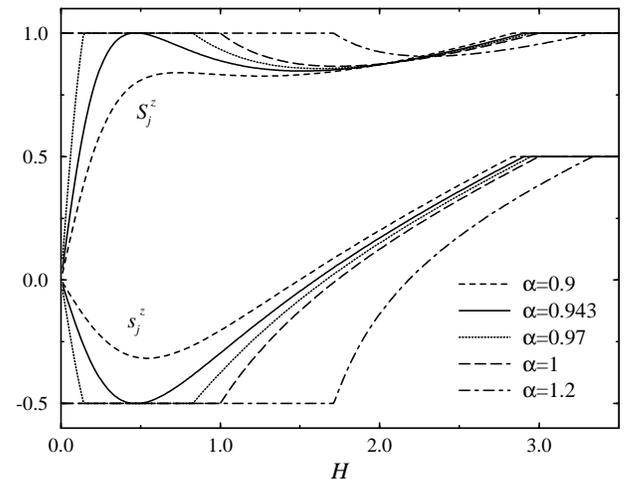}}
\vskip 3mm
\caption{The ground-state spin configurations as functions of the
         field for the classical Hamiltonian (2).}
\label{F:Sclass}
\end{figure}
\begin{figure}
\vskip -5mm
\mbox{\psfig{figure=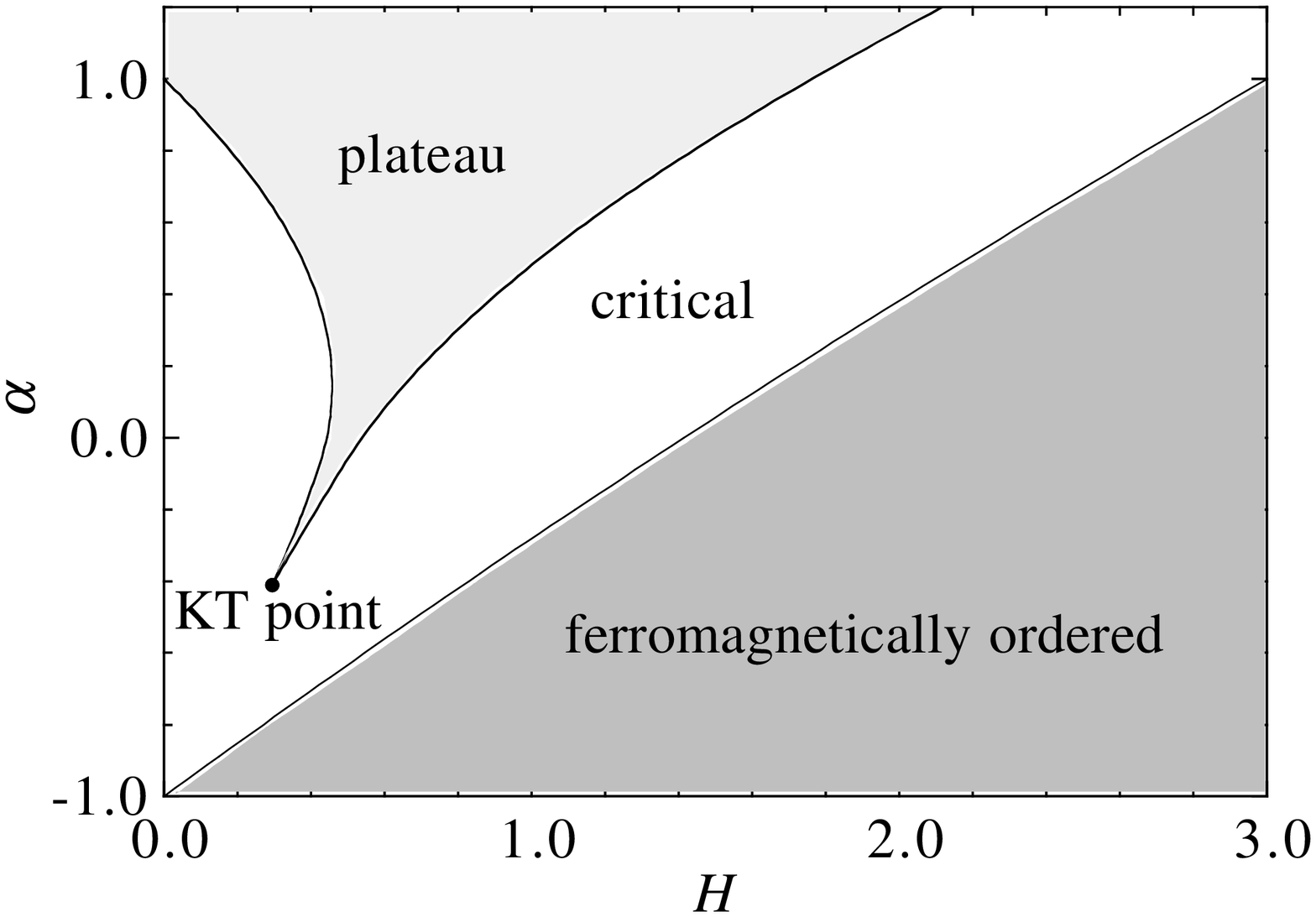,width=80mm,angle=90}}
\vskip 5mm
\caption{The $\alpha$-versus-$H$ phase diagram for the quantum
         Hamiltonian (2).}
\label{F:HaPhD}
\end{figure}

   In order to elucidate how far from intuitive the present
observation is, we compare it with the classical behavior.
Let us consider the classical version of the Hamiltonian (\ref{E:H}),
where $\mbox{\boldmath$S$}_j$ and  $\mbox{\boldmath$s$}_j$ are
classical vectors of magnitude $1$ and $\frac{1}{2}$, respectively.
We show in Fig. \ref{F:mclass} the classical magnetization curves and
learn both similarity and difference between the quantum and
classical behaviors.
In the ferromagnetic and Ising-like antiferromagnetic
exchange-coupling regions they are quite alike, which is convincing
in that quantum effects are supposed to be less significant in both
the regions.
However, the quantum behavior is qualitatively different from the
classical one in the $XY$-like coupling region.
The classical state of $M=N/2$, which is stable enough to form a
plateau in the Ising-like coupling region, is no more massive at
$\alpha\leq 0.943$.
The spin configuration as a function of the field, revealed in Fig.
\ref{F:Sclass}, is suggestive in understanding the prompt collapse of
the classical plateau with the increase of the $XY$-like anisotropy.
In the classical case, the spin configuration in the massive state of
$M=N/2$ is stuck to $(S_j^z,s_j^z)=(1,-\frac{1}{2})$.
In other words, the plateau can not appear unless the configuration
$(S_j^z,s_j^z)=(1,-\frac{1}{2})$ is realized.
This is not the case for the quantum system.
The quantum spin configuration in the massive state of $M=N/2$
generally depends on $\alpha$ and exhibits a quantum reduction from
the classical N\'eel-like state.
At the Heisenberg point, for example, the quantum averages of the
sublattice magnetizations per unit cell in the massive state are
estimated to be $0.793$ and $-0.293$, respectively.
It must be the quantum spin reduction that makes the massive state
tough against the $XY$-like anisotropy.

   In sum the quantum mixed-spin Heisenberg model (\ref{E:H}) with
$(S,s)=(1,\frac{1}{2})$ shows the three distinct phases at the sector
of $M=N/2$;
the plateau phase,
the gapless spin-fluid phase, and
the ferromagnetically ordered phase,
as illustrated in Fig. \ref{F:HaPhD}.
The plateau appears for $\alpha>-0.41$, including a
ferromagnetic-coupling region.
We note that on the boundary of the plateau phase except for the
point $(\alpha,H)=(\alpha_{\rm c},H_{\rm c})\equiv(0.41,0.293)$ which
is indicated as KT point in Fig. \ref{F:HaPhD}, the plateau length
is generally finite, namely, the relevant correlation length is not
divergent.
The only point $(\alpha_{\rm c},H_{\rm c})$ possesses the KT
character.
The long-lived plateau against the $XY$-like anisotropy, which is
contrastive to the corresponding classical behavior, deserves special
remark and further investigation.
We expect magnetic measurements on anisotropic systems \cite{Koni25}.

   It is a pleasure to thank H.-J. Mikeska and U. Schollw\"ock for
helpful discussions.
This work was supported by the Japanese Ministry of Education,
Science, and Culture through Grant-in-Aid No. 09740286 and by the
Okayama Foundation for Science and Technology.
The numerical computation was done in part using the facility of the
Supercomputer Center, Institute for Solid State Physics, University of
Tokyo.

\widetext
\end{document}